\documentclass[twoside,fleqn]{article}
\usepackage{espcrc2}
\usepackage{amssymb}
\usepackage[dvips]{epsfig}
\newcommand{\be}{\begin{equation}}
\newcommand{\ee}{\end{equation}}
\newcommand{\bea}{\begin{eqnarray}}
\newcommand{\eea}{\end{eqnarray}}
\newcommand{\bean}{\begin{eqnarray*}}
\newcommand{\eean}{\end{eqnarray*}}
\newcommand{\T}{\textstyle}

\newcommand{\simas}[1]{\raisebox{-0.1ex}{$\stackrel{\small{#1}}{\sim}$}}

\newcommand{\Fm}{{\rm fm}}

\newcommand{\Or}{\mbox{O}}

\newcommand{\gbar}{\overline{g}}

\newcommand{\lMSbar}{\Lambda_{\overline{{\rm MS}}}}
\newcommand{\Mb}{M_{\rm b}}
\newcommand{\lag}[1]{{\mathcal{L}}_{\rm {#1}}}
\newcommand{\vecsigma}{{\bf \sigma}}
\newcommand{\vecD}{{\bf D}}
\newcommand{\vecB}{{\bf B}}
\newcommand{\heavy}{\psi_{\rm h}}
\newcommand{\heavyb}{\overline{\psi}_{\rm h}}

\newcommand{\lightb}{\overline{\psi}_{\rm l}}
\newcommand{\fa}{f_{\rm A}}
\newcommand{\kv}{k_{\rm V}}
\newcommand{\fone}{f_1}
\newcommand{\fastat}{f_{\rm A}^{\rm stat}}
\newcommand{\fonestat}{f_{1}^{\rm stat}}

\newcommand{\zastat}{Z_{\rm A}^{\rm stat}}
\newcommand{\ZRGI}{Z_{\rm RGI}}
\newcommand{\Xbare}{X_{\rm bare}}
\newcommand{\Xren}{X_{\rm R}}

\newcommand{\XRGI}{X_{\rm RGI}}
\newcommand{\XRGIspin}{X_{\rm RGI}^{\rm spin}}
\newcommand{\Yr}{Y_{\rm PS}}

\newcommand{\Rr}{R_{\rm }}

\newcommand{\gamps}{\Gamma_{\rm PS}}
\newcommand{\gamv}{\Gamma_{\rm V}}
\newcommand{\gamav}{\Gamma_{\rm av}}

\newcommand{\delgam}{{\Delta}_{\Gamma}}
\newcommand{\Cps}{C_{\rm PS}}

\newcommand{\Cspin}{C_{\rm spin}}
\newcommand{\Cpsv}{C_{{\rm PS}/{\rm V}}}
\newcommand{\Cmass}{C_{\rm mass}}
\title{
{
\vspace{-4.5cm} \normalsize \hfill
\parbox{26.0mm}{\raggedleft
MS-TP-04-24\\HU-EP-04/51\\DESY 04-178\\SFB/CPP-04-50\\September 2004}
}\\[25mm]
Non-perturbative tests of HQET in small-volume quenched QCD%
\thanks{Based on a talk presented by J.H. at the conference
LATTICE '04, June 21 -- 26, 2004, Fermilab, Batavia IL, USA.}%
\thanks{Work supported by the DFG in the SFB/TR 09.}
}
\author{
Jochen Heitger\address{
Universit\"at M\"unster, Institut f\"ur Theoretische Physik,
D-48149 M\"unster, Germany},
Andreas J\"uttner\address{
Humboldt Universit\"at Berlin, Institut f\"ur Physik,
D-12489 Berlin, Germany},
Rainer Sommer\address{
DESY, D-15738 Zeuthen, Germany, and CERN, Theory Division, 
CH-1211 Geneva 23, Switzerland} and
Jan Wennekers\address{
DESY, Theory Group, D-22603 Hamburg, Germany}
(ALPHA Collaboration)
}
\begin{document}
%
\makeatletter
\long\def\@maketablecaption#1#2{\vskip 10mm #1. #2\par}
\makeatother
%
\begin{abstract}
We quantitatively investigate the quark mass dependence of current 
matrix elements and energies, calculated over a wide range of quark 
masses in the continuum limit of small-volume quenched lattice QCD.
By a precise comparison of these observables as functions of the heavy 
quark mass with the predictions of HQET we are able to verify that their 
large quark mass behaviour is described by the effective theory.
\end{abstract}
\maketitle
%
%
\section{MOTIVATION}
The simplification of the QCD dynamics in the limit of large masses of 
c-- and b--quarks gives rise to the Heavy Quark Effective Theory (HQET) 
as a standard phenomenological tool to describe decays of heavy-light 
hadrons and their transitions in terms of hadronic matrix elements. 
(See e.g.~\cite{HQET:neubert} for a review.)
Starting from the HQET Lagrangian of a heavy quark, $\lag{HQET}$, that
reads
\be
\heavyb\Big[
D_{0}+m-{\T \frac{\omega_{\rm kin}}{2m}}\vecD^2 
-{\T \frac{\omega_{\rm spin}}{2m}}\vecsigma\cdot\vecB 
\Big]\heavy+\Or({\T \frac{1}{m^2}})\,,\,
\label{lag}
\ee
this effective theory provides an expansion of the QCD amplitudes in the
inverse heavy quark mass, $1/m$, and is renormalizable at any finite 
order in $1/m$ according to power counting.

Making HQET an effective theory for QCD requires matching calculations 
to express the parameters $m,\omega_{\rm spin},\ldots$ in (\ref{lag}) by 
those of QCD, and the agreement of different determinations of 
quantities such as $V_{\rm cb}$ \cite{reviews:Stone03}, which involve
perturbative HQET, already reflect the success of the effective theory 
approach.
Apart from these phenomenological tests of HQET, however, independent, 
\emph{non}-perturbative ones are still desirable.

Though in principle the lattice achieves this by varying $m$, clean 
comparisons of QCD and HQET in the \emph{continuum limit} demand 
$m\ll 1/a$ prior to $a\to 0$.
Yet \cite{HQET:pap1} rather turned that restriction into an idea to 
connect them non-perturbatively:
consider QCD in a small volume $L^4$, where the ${\rm b}$ can be 
simulated as relativistic fermion and HQET becomes an expansion of QCD 
in the variable
\[
1/z \equiv 1/(ML)\,,\,\,\,
\mbox{$M$: RGI heavy quark mass}\,.
\]

Here, we confront the large--$z$ behaviour of correlation functions, 
computed with Schr\"odinger Functional boundary 
conditions \cite{SF:LNWWS}, to the static theory and estimate the size 
of $1/m$--corrections.
This is also of practical relevance in the strategy to solve 
renormalization problems in HQET non-perturbatively by matching to QCD 
in finite volume \cite{HQET:pap1}.
For a full report on our study (and a list of references to related 
work) consult \cite{HQET:pap3}.
\section{LARGE-MASS ASYMPTOTICS}
Our observables derive from relativistic heavy-light SF correlation 
functions.
With $\fa$ a correlator between a heavy-light pseudoscalar boundary 
source and the axial current 
$A_0=\lightb\gamma_{0}\gamma_5\psi_{\rm b}$ in the bulk, $\kv$ its 
vector channel analogue and $\fone$ a boundary-to-boundary correlation, 
we define
\be\T
\Yr(L,M) \equiv \frac{\fa(T/2)}{\sqrt{\fone}}\,,\,\,\,
\Rr(L,M) \equiv -\frac{\fa(T/2)}{\kv(T/2)}\,, 
\label{ratios}
\ee
which from the multiplicative renormalizability of the SF boundary 
fields follow to be finite quantities provided that $A_0,V_0$ denote 
renormalized currents.
Effective energies are constructed as
\[
\gamps(L,M)\equiv
-\fa'(T/2)/\fa(T/2)\,,\,\,\,
\mbox{$\gamv$:$\,\fa\to\kv$}
\]
with spin-averaged sum and difference
\be
\gamav\equiv{\T \frac{1}{4}}\left[\,\gamps+3\,\gamv\,\right]\,,\,\,\,
\delgam\equiv\gamps-\gamv\,.
\label{sumdiff}
\ee
Being (ratios of) matrix elements between low-energy heavy-light and 
vacuum-like states \cite{HQET:pap3}, (\ref{ratios}) and (\ref{sumdiff}) 
should then be described by HQET, which we test by verifying their 
large--$z$ asymptotics to comply with the predictions of HQET.

To obtain them, note that classically one expects the current matrix 
elements to be power series in $1/z$ led by the \emph{static limit}, 
where the heavy quark does not propagate in space.
E.g., replacing in (\ref{ratios}) $\psi_{\rm b}$ by $\heavy$ and 
dropping the $\Or(1/m)$ of $\lag{HQET}$ in (\ref{lag}) to evaluate the 
static correlators, the effective theory and QCD are related by
\be
X(L)\equiv
\fastat({\T \frac{T}{2}})/{\T \sqrt{\fonestat}}=
\lim_{z\to\infty}\Yr(L,M)\,.
\ee
On the quantum level, the scale dependent renormalization of the 
effective theory implies logarithmic modifications, i.e.~the axial 
current renormalization amounts $\Xren(L,\mu)=\zastat(\mu)\Xbare(L)$ to 
depend logarithmically on the chosen renormalization scale ($\mu$) and 
scheme, but not so the associated \emph{renormalization group invariant} 
\be 
\XRGI(L)= 
\lim_{\mu\to\infty}\Big\{
[2b_0\gbar^2(\mu)]^{-\frac{\gamma_0}{2b_0}}\Xren(L,\mu)\Big\}\,,\, 
\label{XRGI}
\ee
calculable in lattice QCD: $\XRGI=\ZRGI\Xbare$ \cite{zastat:pap3}.
The large--$z$ behaviour of (\ref{ratios}) thus splits into RGIs of the 
effective theory and logarithmically mass dependent functions, $C$.
As their arguments we choose $r\equiv M/\lMSbar$, since it can be fixed 
on the lattice without perturbative 
uncertainties \cite{msbar:pap1,HQET:pap2}:
\bea
\Yr(L,M)
& \hspace{-2.0ex}\simas{M\to\infty}\hspace{-1.75ex} &
\Cps(r)\,\XRGI(L)+\Or(1/z)\,,
\label{yR2stat}\\
\Rr(L,M) 
& \hspace{-2.0ex}\simas{M\to\infty}\hspace{-1.75ex} &
\Cpsv(r)\left(1+\Or(1/z)\right)\,.
\label{rR2stat}
\eea
Similar predictions hold for the energies (\ref{sumdiff}),
\bea
L\gamav(L,M)
& \hspace{-2.0ex}\simas{M\to\infty}\hspace{-1.75ex} &
\Cmass(r)\,z+\Or({\T \frac{1}{z^0}})\,,
\label{gamasympt}\\
L\delgam(L,M)
& \hspace{-2.0ex}\simas{M\to\infty}\hspace{-1.75ex} &
\Cspin(r)\,\frac{\XRGIspin(L)}{z}+\Or({\T \frac{1}{z^2}})\,,\,\,
\label{delgamasympt}
\eea
where $\Cmass$ translates the pole into the RGI quark mass and 
$\XRGIspin$ is a RGI matrix element of the chromomagnetic operator 
$\heavyb\vecsigma\hspace{-0.5ex}\cdot\hspace{-0.5ex}\vecB\heavy$, whose 
anomalous dimension (AD) contributes to $\Cspin$.

Taking the entering ADs to best perturbative 
accuracy \cite{HQET:pertADs}, the conversion functions $C$ in 
(\ref{yR2stat})--(\ref{delgamasympt}) are evaluated by solving 
RGEs \cite{HQET:pap3}. 
As detailed there, their perturbative knowledge is adequate to allow for 
an investigation of the $1/z^n$--corrections.
\section{DISCUSSION OF THE RESULTS}
\begin{figure}[t]
\centering
\vspace{-1.125cm}
\epsfig{file=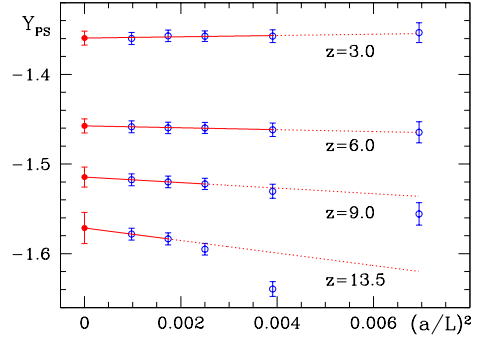,width=7.0cm}
\vspace{-0.25cm}
\caption{{\sl
Continuum limit extrapolations spanning the entire $z$--range.
(Dotted lines extend the linear fits in $(a/L)^2$ to omitted values of 
$a/L$.)
}}\label{fig:yR_CL}
\vspace{-0.5cm}
\end{figure}

Our quenched data refer to a volume of extent $L=T\approx 0.2\,\Fm$, 
which admits to reach $M>2\Mb$ while $a\to 0$ extrapolations are still 
well controllable.
To account for the growing of the quark mass in lattice units at given
$a/L$ as $z$ is increased, the coarsest resolutions that pass into the 
continuum extrapolations linear in $(a/L)^2$ are chosen by imposing a 
cut on $aM$ \cite{HQET:pap3,HQET:pap2}.
Fig.~\ref{fig:yR_CL} illustrates a typical case:
the slopes are quite small, and the error in the continuum limit gets 
the larger the more $a/L$ are to be discarded for increasing $z$.

The main results and their polynomial fits in $1/z$ to quantify the 
deviations from the static limit are displayed in 
figs.~\ref{fig:yR_z}--\ref{fig:delgam_z}; linear fits use only the 
heavier quark mass points.
\subsection{Current matrix elements}
\begin{figure}[t]
\centering
\vspace{-1.0cm}
\epsfig{file=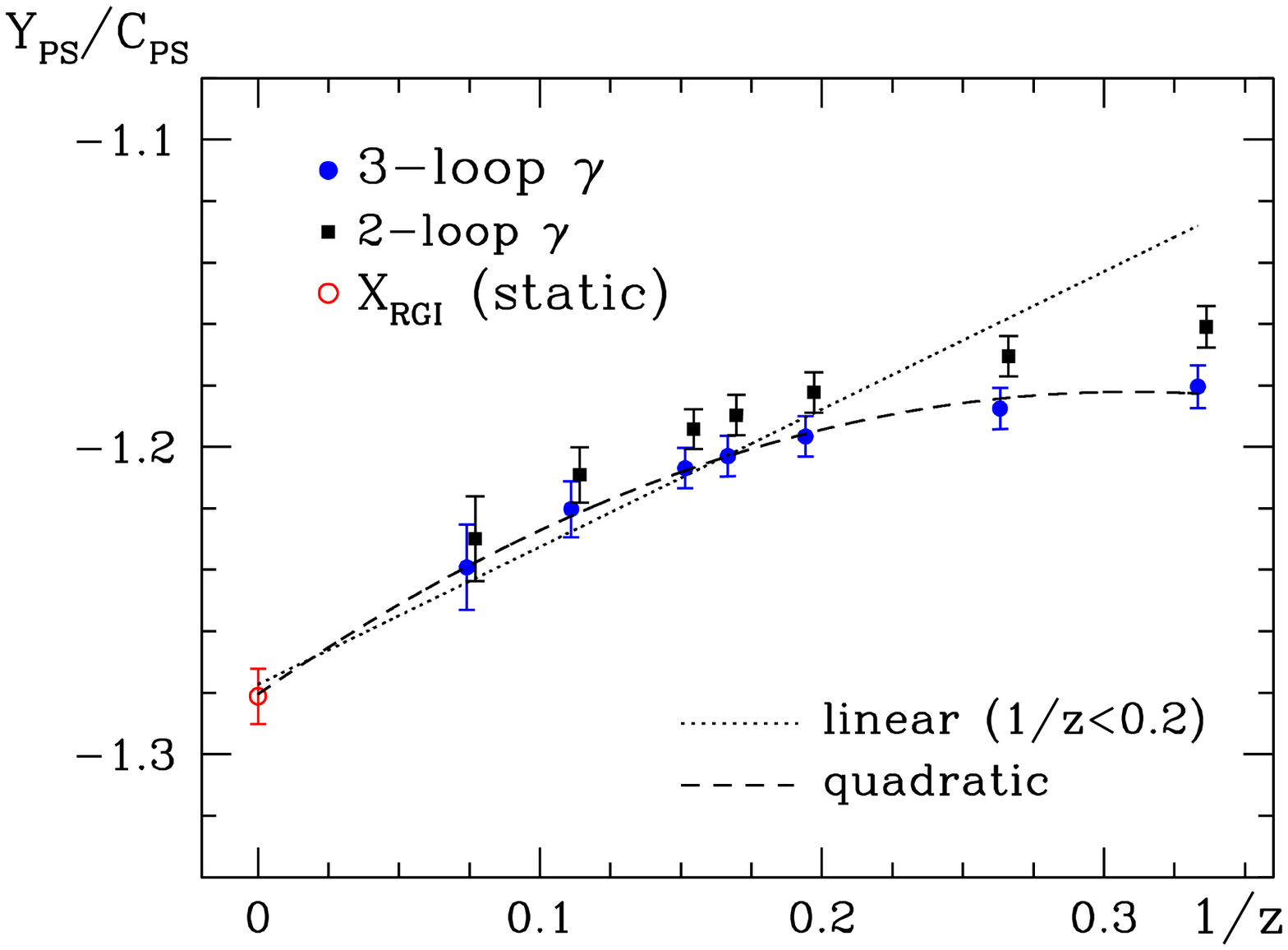,width=7.0cm}
\vspace{-0.25cm}
\caption{{\sl
Fits of $\Yr/\Cps$ include $\XRGI$.
The AD of the static axial current ($\gamma$) enters in $\Cps$.
}}\label{fig:yR_z}
\end{figure}

\begin{figure}[t]
\centering
\vspace{-1.25cm}
\epsfig{file=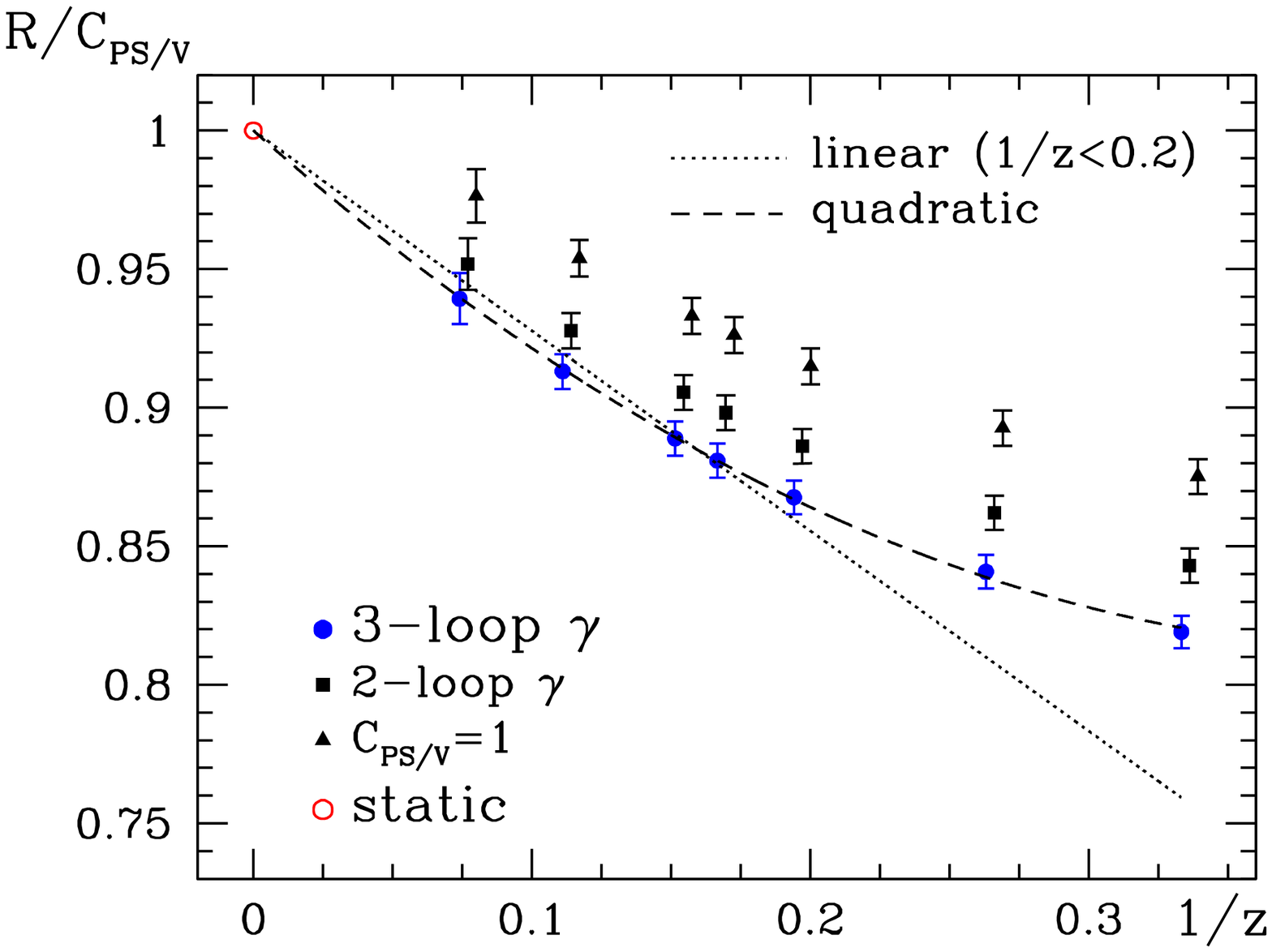,width=7.0cm}
\vspace{-0.25cm}
\caption{{\sl
Fits of $\Rr/\Cpsv$, constrained to 1.
}}\label{fig:rR_z}
\vspace{-0.5cm}
\end{figure}

Comparing the finite-mass matrix element of $A_0$ with the HQET 
prediction $\XRGI(L)+\Or(1/z)$, we infer from  fig.~\ref{fig:yR_z} that 
the perturbative $\Cps$ reduces the mass dependence of $\Yr$ 
significantly and renders $\Yr(L,M)/\Cps(r)$ to be compatible with 
approaching the static result for $\XRGI$ \cite{HQET:pap3,zastat:pap3} 
as $1/z\to 0$.
Also the ratio $\Rr$ of matrix elements of $A_0$ and $V_0$ in 
fig.~\ref{fig:rR_z} is consistent with the leading term in the 
$1/z$--expansion (fixed to 1 by the spin symmetry of the static theory, 
cf.~(\ref{rR2stat})), if $\Cpsv$ is evaluated including at least the 
current's two-loop ADs.
In both cases, the coefficients of $1/z$--corrections are of order one 
and, therefore, the corrections are reasonably small.
\subsection{Effective energies}
\begin{figure}[t]
\centering
\vspace{-1.0cm}
\epsfig{file=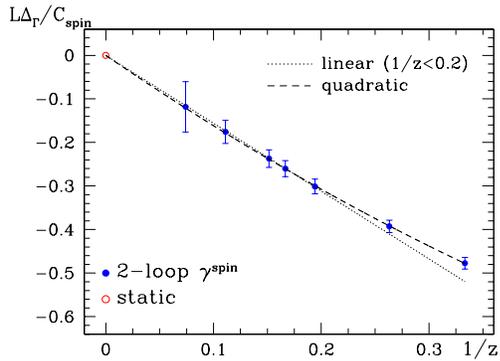,width=7.0cm}
\vspace{-0.25cm}
\caption{{\sl
Constrained fits with $\Cspin$ to 2 loops.
}}\label{fig:delgam_z}
\vspace{-0.875cm}
\end{figure}

In confirming the asymptotics (\ref{gamasympt}), the smallness of 
$(1/z)^2$--terms found in the combination $L\gamav/(z\Cmass)$ deserves 
particular emphasis regarding the static limit computation of the b-mass 
via $\gamav$ \cite{HQET:pap1,HQET:pap2}, since it yields an estimate of 
the error to $M_{\rm b}$, originating from the matching to QCD, of only 
$\approx 1\%$ \cite{HQET:pap3}.
The spin splitting $L\delgam/\Cspin$ (fig.~\ref{fig:delgam_z}) vanishes 
for $1/z\to 0$ as expected, exhibiting a rather large 
$1/z$--coefficient.
\subsection{Summary}
Our successful tests of HQET show the \emph{continuum limits} of the 
\emph{non-perturbatively renormalized} QCD observables at finite $z$ to
meet the predictions of the effective theory.
Only the functions $C$ relating them to the RGIs of the latter induce
perturbative uncertainties, but these are under control (except for 
$\Cspin$ lacking the NNLO) and reveal the power corrections to dominate 
over perturbative ones in the accessible $z$--range.
Finally, our results appear promising to determine 
$1/m_{\rm b}$--corrections to B-physics matrix elements 
following \cite{HQET:pap1} by extending the non-perturbative matching of 
HQET and QCD to subleading terms \cite{HQET:pap3,lat04:stephan}.
%
%
%

%
\end{document}